\documentclass[5p,a4paper]{elsarticle}
\journal{Superconductivity}

\usepackage[utf8]{inputenc}
\usepackage[T1]{fontenc}
\usepackage{microtype,booktabs,afterpage}

\biboptions{sort&compress}
\usepackage{lineno}
\modulolinenumbers[5]

\usepackage{amsmath,bm}

\usepackage[pdftex]{color}
\usepackage{verbatim}
\pdfoutput=1

\usepackage[colorlinks,plainpages=false,linkcolor=blue,urlcolor=blue,citecolor=blue,pdfpagemode=UseNone,pdfstartview=FitBH]{hyperref}
\usepackage[charter,greekuppercase=italicized]{mathdesign}

\usepackage{graphicx}    
\graphicspath{{./}{figures/}}
\DeclareGraphicsExtensions{.eps,.png,.pdf,.jpg}

\definecolor{red}{rgb}{0.85,.1,0}
\definecolor{green}{rgb}{0.0,0.8,0.0}
\definecolor{orange}{rgb}{1,0.5,0}

\SetExpansion[context = sloppy,shrink = 60]{encoding = {OT1,T1,TS1} }{}

\begin{document}

\begin{frontmatter}

\makeatletter\renewcommand{\ps@plain}{%
\def\@evenhead{\hfill\itshape\rightmark}%
\def\@oddhead{\itshape\leftmark\hfill}%
\renewcommand{\@evenfoot}{\hfill\small{--~\thepage~--}\hfill}%
\renewcommand{\@oddfoot}{\hfill\small{--~\thepage~--}\hfill}%
}\makeatother\pagestyle{plain}


\title{Single-gap two-band superconductivity well above the Pauli limit in non-centrosymmetric TaIr\textsubscript{2}B\textsubscript{2}}
\author[1]{J.~Ka\v{c}mar\v{c}\'ik}
\author[1]{Z.~Pribulov\'a}
\author[2,3]{T.~Shiroka\corref{cor1}}
\ead{tshiroka@phys.ethz.ch}
\author[1,4]{F.~Ko\v{s}uth}
\author[1]{P.~Szab\'o}
\author[5]{M.~J.~Winiarski}
\author[5]{S.~Kr\'olak}
\author[6]{J.~Jaroszynski}
\author[7]{T.~Shang}
\author[8]{R.~J.~Cava}
\author[9]{C.~Marcenat}
\author[10]{T.~Klein}
\author[5]{T.~Klimczuk\corref{cor2}}
\ead{tomasz.klimczuk@pg.edu.pl}
\author[1]{P.~Samuely}
\address[1]{Centre of Low Temperature Physics, Institute of Experimental Physics, Slovak Academy of Sciences, SK‑04001 Košice, Slovakia}
\address[2]{Laboratorium f\"{u}r Festk\"{o}rperphysik, ETH Zürich, 8093 Zurich, Switzerland}
\address[3]{PSI Center for Neutron and Muon Sciences CNM, 5232 Villigen PSI,  Switzerland }
\address[4]{Centre of Low Temperature Physics, Faculty of Science, P. J. Šafárik University, SK-04001 Košice, Slovakia}
\address[5]{Faculty of Applied Physics and Mathematics and Advanced Material Center, Gdansk University of Technology, ul. Narutowicza 11/12, Gdańsk 80–233, Poland}
\address[6]{National High Magnetic Field Laboratory, Florida State University, Tallahassee, Florida 32310, USA}
\address[7]{Key Laboratory of Polar Materials and Devices (MOE), School of Physics and Electronic Science, East China Normal University, Shanghai 200241, China}
\address[8]{Department of Chemistry, Princeton University, Princeton, 08544, New Jersey, USA}
\address[9]{Univ. Grenoble Alpes, CEA, Grenoble INP, IRIG, PHELIQS, 38000, Grenoble, France}
\address[10]{Univ. Grenoble Alpes, CNRS, Institut Néel, 38000, Grenoble, France}

\cortext[cor1]{Corresponding author}
\cortext[cor2]{Corresponding author}

\date{\today}

\begin{abstract}
Non-centrosymmetric superconducting materials represent an exciting class
of novel superconductors featuring a variety of unconventional properties,
including mixed-parity pairing and very high upper critical fields. Here,
we present a comprehensive study of TaIr$_2$B$_2$ (with $T_c = 5.1$\,K),
using a set of complementary experimental methods, including bulk- and
surface-sensitive techniques. 
We provide evidence that this system is a two-band, yet it behaves as
a single-gap superconductor with a strong coupling. The upper critical
field of TaIr$_2$B$_2$ significantly exceeds the Pauli limit and
exhibits a nearly linear temperature dependence down to the lowest
temperatures. This behavior, rarely seen in superconductors, is
discussed in terms of anti-symmetric spin-orbit interaction, two-band-,
and strong-coupling effects, as well as disorder.
\end{abstract}

\begin{keyword}
Non-centrosymmetric superconductors \sep
Multiband/multigap superconductivity \sep
Spin-orbit coupling \sep
Upper critical magnetic field \sep
Pauli limit \sep
Muon-spin spectroscopy \sep
STM spectroscopy
\end{keyword}

\end{frontmatter}


\section{\label{sec:intro}Introduction}

Together with the transition temperature $T_c$, the upper critical field
$B_{c2}$, which suppresses the Cooper pairs, is one of the fundamental
properties of the superconducting state. In most cases, the magnetic
field acts first on the orbital motion of the Cooper pairs via the
Lorentz force exerted on them, and Abrikosov vortices with normal cores
are introduced. When these cores overlap, the $B_{c2}$ limit is reached.
However, this mechanism becomes inefficient in case of spatial confinement.
In such cases, the Pauli limit, where pair breaking occurs via spin
splitting, represents the ultimate upper critical field.
This happens when the Zeeman energy is equal to the superconducting
energy gap. In non-centrosymmetric (NCS) systems this is not necessarily
the case. Without the inversion symmetry an antisymmetric spin-orbit
coupling lifts the Kramers degeneracy and leads to spin-split bands~\cite{Gorkov2001,Singh2018}, which can feature unconventional superconductivity
with mixed parity. In the Ce-based heavy-fermion superconductors~\cite{Bauer2004,Bauer2012} the paramagnetic pair breaking effects are suppressed due to the Rashba
type spin-orbit coupling, resulting in a mixture of singlet- and triplet
pairing. Such mixed-parity superconductors can feature upper critical
fields significantly larger than the Pauli limit, 
$B_\mathrm{P} = \Delta/(\sqrt{g}\mu_\mathrm{B}$), with $\Delta$ the
superconducting energy gap, $g$ the Land\'e $g$-factor, and $\mu_\mathrm{B}$
the Bohr magneton. Here, the orbital upper critical field is also enhanced
due to the to small coherence length arising from the heavy fermions.
In atomically thin transition-metal dichalcogenide films, like the NbSe$_2$ monolayer~\cite{Xi2016}, as well as in quasi-2D bulk crystals, such as
(LaSe)$_{1.14}$NbSe$_2$ misfit compounds~\cite{Samuely2023}, the
orbital pair breaking is ineffective within the planes due to the short
interlayer coherence length. In such cases, extremely large in-plane
upper critical fields, much above the theoretical Pauli limit, are found,
mostly arising from the Ising spin-orbit coupling (SOC) allowed by
the broken inversion symmetry. The Ising SOC locks the orientation of
the electron spins to the out-of-plane direction and makes the action of
the in-plane external magnetic field highly irrelevant. Ising spin–orbit coupling also applies to the quasi-1D material K$_2$Cr$_3$As$_3$~\cite{Bao2015,Smidman2017}. Two other reasons may also enhance the upper critical field in NCS: a reduced Land\'e $g$-factor due to spin-orbit coupling and the strong-coupling effects with $\Delta > \Delta_\mathrm{BCS}$ or multiple gaps. 

\begin{figure}[thb]
\centering
\vspace{-2mm}
\includegraphics[width=0.37\textwidth]{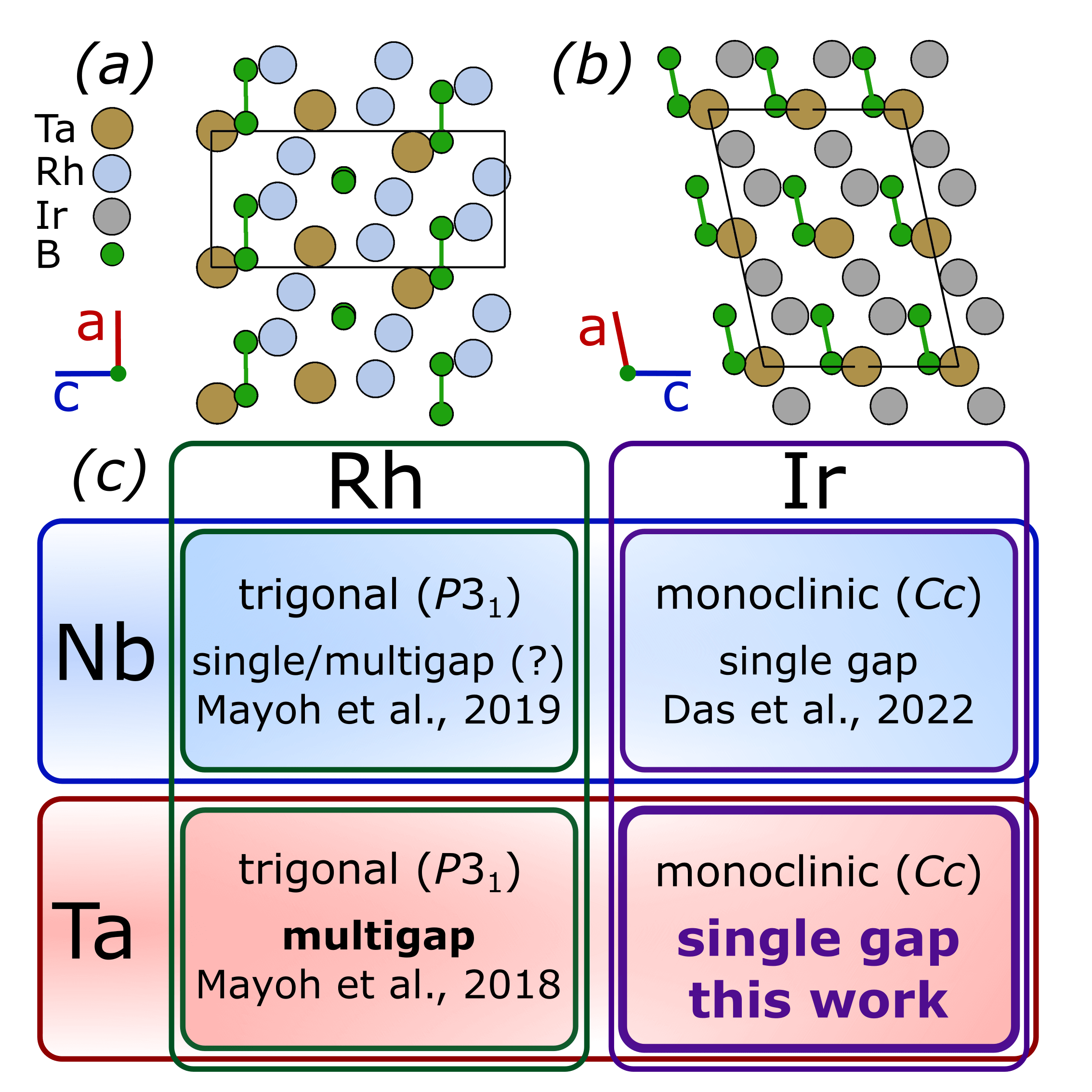}
\caption{a) Trigonal crystal structure of (Nb/Ta)Rh$_2$B$_2$, with space group $P3_1$ no.\ 144 and b) monoclinic structure of (Nb/Ta)Ir$_2$B$_2$, with space group $Cc$ no.\ 9. c) Summary of studies on the superconducting gap of (Nb/Ta)(Rh/Ir)$_2$B$_2$.}
\label{fig:structure}
\end{figure}

Recently, four new ternary superconductors with a low symmetry non-centrosymmetric crystal structure have been found. Namely, NbRh$_2$B$_2$ and TaRh$_2$B$_2$, sharing the chiral space group $P3_1$~\cite{Carnicom2018}, with superconducting $T_c = 7.6$\,K and 5.8\,K, respectively (Fig.~\ref{fig:structure}a) and the isoelectronic compounds  NbIr$_2$B$_2$ and TaIr$_2$B$_2$, with a monoclinic $Cc$ space group
and $T_c = 7.2$\,K and 5.1\,K~\cite{Gornicka2021}, respectively (Fig.~\ref{fig:structure}b). Their structural similarities and small differences are discussed in Ref.~\cite{Gornicka2021}. First-principle calculations highlight the key role of the antisymmetric spin-orbit coupling, which
leads to a splitting of the Fermi surface sheets in all four compounds.
While all of them feature high upper critical fields $B_{c2}$, well above
the Pauli limit $B_\mathrm{P}$, the strongest effects are found in TaIr$_2$B$_2$. In NbRh$_2$B$_2$ and TaRh$_2$B$_2$ indications of multiple superconducting gaps are found~\cite{Mayoh2018,Mayoh2019}, while in NbIr$_2$B$_2$ a single-gap
scenario was reported [see (Fig.~\ref{fig:structure}c)]. Interestingly,
based on \emph{ab initio} calculations, (Nb/Ta)Ir$_2$B$_2$ were classified as Weyl semimetals in the normal state~\cite{Gao2021}.

Here, we focus on a detailed study of the TaIr$_2$B$_2$ superconductor,
the heaviest compound of all four systems, where SOC effects should be
the strongest. By using tunneling- and {\textmu}SR spectroscopy, as well
as highly sensitive AC calorimetry, we show that TaIr$_2$B$_2$ has an
$s$-wave superconducting order parameter, it shows a single gap, and
exhibits a strong electron-phonon coupling. On the other hand, the field
dependence of the {\textmu}SR relaxation rate can only be explained by
a two-band model, indicating that TaIr$_2$B$_2$ is a single-gap but
two-band superconductor. Consistent with the strongest spin-orbit coupling,
this compound has the highest $B_{c2}$/$B_\mathrm{P}$ ratio, with an
experimentally established $B_{c2} = 17.4$\,T (from calorimetry and
transport measurements) that is almost 50\% higher than the Pauli field
$B_\mathrm{P}$ for strong coupling.
This is a significantly stronger effect than in the other three compounds. 
$B_{c2}$ is linear in temperature down to the lowest temperature (here,
300\,mK), which is very unusual. Our findings are discussed in terms of
anti-symmetric spin-orbit interaction, two-band effects, strong coupling,
and disorder.

\section{Experimental part}

\subsection{Synthesis}
TaIr$_2$B$_2$ was synthesized by using a one-step solid-state reaction
technique~\cite{Gornicka2021}. The starting materials were elemental
tantalum (99.9\% pure, 100 mesh, Alfa Aesar), iridium (99.99\%,
Mennica-Metale, Poland) and boron (submicron particles, Callery Chemical).
Powders of Ta, Ir and B were weighted out in 1:1.98:2.15 ratio, ground
very finely using a mortar and agate pestle and then pressed into a
pellet using a hydraulic press. The ratio of elements was modified
compared to our first report~\cite{Gornicka2021} (1:2:2.33), resulting
in a better phase purity. The sample (with a mass of $\sim 180$\,mg) was
then wrapped in a tantalum foil, placed in an alumina crucible, and
heat-treated at 1070$^{\circ}$C for 6 hours under dynamic high vacuum
($10^{-6}$\,mbar). There was no significant weight loss observed after
the synthesis. 
The crystal structure and the phase composition were evaluated by means
of powder X-ray diffraction (pXRD), using a Bruker D2 Phaser diffractometer
with Cu K$\alpha$ source and a Lynxeye XT-E position-sensitive solid-state
detector. A trace amount of elemental Ir was observed in the sample.
The powder X-ray diffraction pattern of TaIr$_2$B$_2$ is shown in Fig.~S1
of the Supplemental Material~\cite{SM}. The refined atomic coordinates
and the isotropic displacement parameters of TaIr$_2$B$_2$ are listed
in Table~S1 of Supplemental Material~\cite{SM} as well.

\subsection{Transport}
Electrical resistivity measurements were performed using the four-probe
method employing a Quantum Design Evercool-II Physical Properties Measurement
System (PPMS). Thin platinum wires were attached to a polished flat
sample using a two-part conductive silver epoxy. Measurements of the
superconducting transition were performed in applied magnetic fields up
to $\mu_0H = 9$\,T. High field transport measurements were performed at
the National High Magnetic Field Laboratory at the Florida State University
in the 35\,T resistive magnet down to 0.3\,K. The resistive transitions
at selected temperatures are depicted in Fig.~S3 of the Supplemental
Material~\cite{SM}.

\subsection{STM/STS}
To study the quasiparticle density of states (DOS) of TaIr$_2$B$_2$, we
carried out low-temperature scanning tunneling micro\-sco\-py/\-spec\-tro\-scopy
(STM/STS) measurements. The experiments were performed in a home-made STM
system immersed in a Janis SSV $^3$He cryostat, which allows measurements
down to $T = 0.3$\,K in magnetic fields up to $B = 8$\,T. The measured
surfaces of the investigated samples were prepared by cleavage under
ambient conditions. The as-prepared sample was mounted into the head of
the STM system and then cooled down to 0.4\,K.
The Au needle was used to form a tunneling junction. The symmetry and
the value of the superconducting energy gap were determined from fitting
the tunneling spectra to the thermally smeared Dynes modification of the
BCS DOS, taking into account the spectral broadening $\Gamma$ in the form
of a complex energy, i.e., d$I$/d$V$ $N(E)$ = Re{$\frac{E-i\Gamma}{\sqrt{(E-i\Gamma)^2-\Delta^2}}$}, where $\Delta$ is the superconducting energy gap. The temperature-
and magnetic field dependencies of the tunneling spectra were measured at
different fixed values of temperatures and magnetic fields, respectively.

\subsection{{\textmu}SR}
To study the microscopic properties of TaIr$_2$B$_2$ in its superconducting
state, we performed transverse-field (TF) {\textmu}SR experiments on the
GPS and Dolly spectrometers of the S{\textmu}S facility at the Paul Scherrer
Institute (Villigen, Switzerland). Both instruments employ a 28\,MeV/c surface
muon beam with the decay positrons being detected by a system of four
detectors. In our case, the forward and backward positron detectors (with
respect to the initial muon polarization) were used to follow the evolution
in time of the {\textmu}SR asymmetry, $A(t)$. The sample consisted of a pellet,
5\,mm in diameter and 1\,mm thick. The fraction of muons stopping in the
sample was approximately 80\% of the incoming muons, the rest passing
undetected through the thin sample holder. The measurements in the
superconducting state were performed as a function of increasing temperature
after cooling the sample in an applied magnetic field. This procedure allows
a regular penetration of magnetic flux (i.e., it avoids issues due to
flux pinning), thus allowing us to probe the intrinsic superconducting
properties. We also performed measurements at base temperature as a function
of the applied field. Again, at each field change, we warmed up the sample
above $T_{c}$, before cooling it in an applied field.

\subsection{Heat capacity}
AC heat capacity measurements were performed in a static magnetic field
(0--8\,T in Košice, and 7--19\,T in Grenoble) using a resistive chip.
The sample was attached to the backside of a bare Cernox resistive chip
using a small amount of Apiezon grease. The resistive chip was split
into heater and thermometer parts by artificially making a notch along
the middle line of the chip. The heater part was used to generate a
periodically modulated heating power $P_{ac}$. The induced oscillating
temperature $T_{ac}$ of the sample was monitored by the thermometer part
of the resistive chip. For details about the method see Ref.~\cite{Yang}.
The AC heat capacity measurements at 0\,T and at 10\,T were also performed
using a chromel-constantan thermocouple serving as a thermometer and as
a sample holder at the same time, while the heat was supplied to the
sample from an LED via an optical fiber, as described elsewhere~\cite{Sullivan}.
Such configuration significantly reduces the addenda contribution to
the total measured heat capacity and enables measurements of the sample's
heat capacity with a high resolution. However, the resulting data is in
arbitrary units. In both experimental approaches, the precise calibrations
and corrections in magnetic fields were included in the measurements
and the data treatment.

\begin{figure*}[ht]
\centering
\vspace{-2mm}
\includegraphics[width=1.0\textwidth]{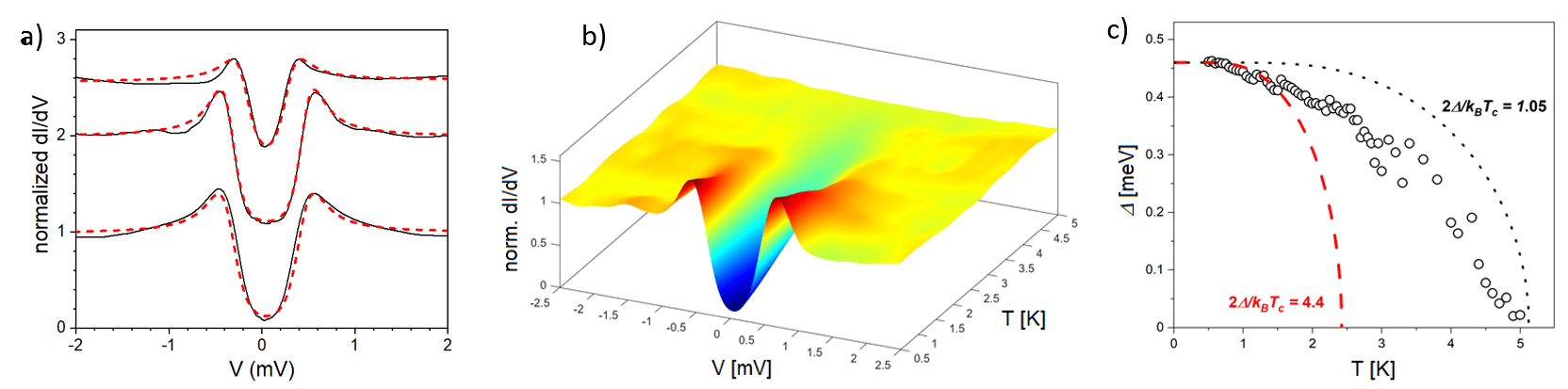}
\caption{a) Typical STM spectra measured on TaIr$_2$B$_2$ surfaces
(solid lines) at $T = 0.5$\,K shifted along the $y$-axis for clarity,
plotted together with the fitting curves to the thermally smeared $s$-wave,
one-gap BCS DOS (red dashed lines). The local values of the superconducting
energy gap $\Delta(0)$, determined from fitting are 0.25\,meV, 0.4\,meV
and 0.45\,meV for the curves at the top, middle, and bottom, respectively.
b) 3D plot of the temperature dependence of the tunneling spectra with
$\Delta(0) = 0.45$\,meV. c) temperature dependence of the superconducting
energy gap determined from fitting to BCS DOS (black symbols). The black
dotted and red dashed lines plot BCS like $\Delta$(T) dependencies with
$\Delta$(0) = 0.46 meV and $T_c$'s 2.4\,K and 5.1\,K, respectively. The
corresponding coupling strengths $2\Delta/k_\mathrm{B}T_c$ are 1.05 and 4.4.}
\label{fig:STM1}
\end{figure*}

\section{Results}

The pXRD pattern of TaIr$_2$B$_2$ (see Fig.~S1 in the Supplementary Material)
is consistent with the previously reported non-centrosymmetric $Cc$
cell~\cite{Gornicka2021}. In a recent report, Lozanov et al.~\cite{lozanov_crystal_2024} suggested that TaIr$_2$B$_2$ crystallizes in a triclinic centrosymmetric
(s.g.\ $P\overline{1}$) structure. However, their claim is based on results
of density functional theory (DFT) calculation of the total energy performed
without relaxing the cell parameters, while the presented nuclear magnetic
resonance (NMR) spectra are instead consistent with the monoclinic $Cc$
crystal structure. Further, our preliminary DFT results suggest that the
fully relaxed $Cc$ cell is energetically favorable over the $P\overline{1}$
cell. Since the samples were fine-grounded and pressed into a pellet, we
used a scanning electron microscope to visualize the particle-size
distribution. The grains were found to have a typical size of approximately
5\,{\textmu}m (see Fig.~S2 in Supplemental Material~\cite{SM}).

\begin{figure}[th]
\centering
\includegraphics[width=0.36\textwidth]{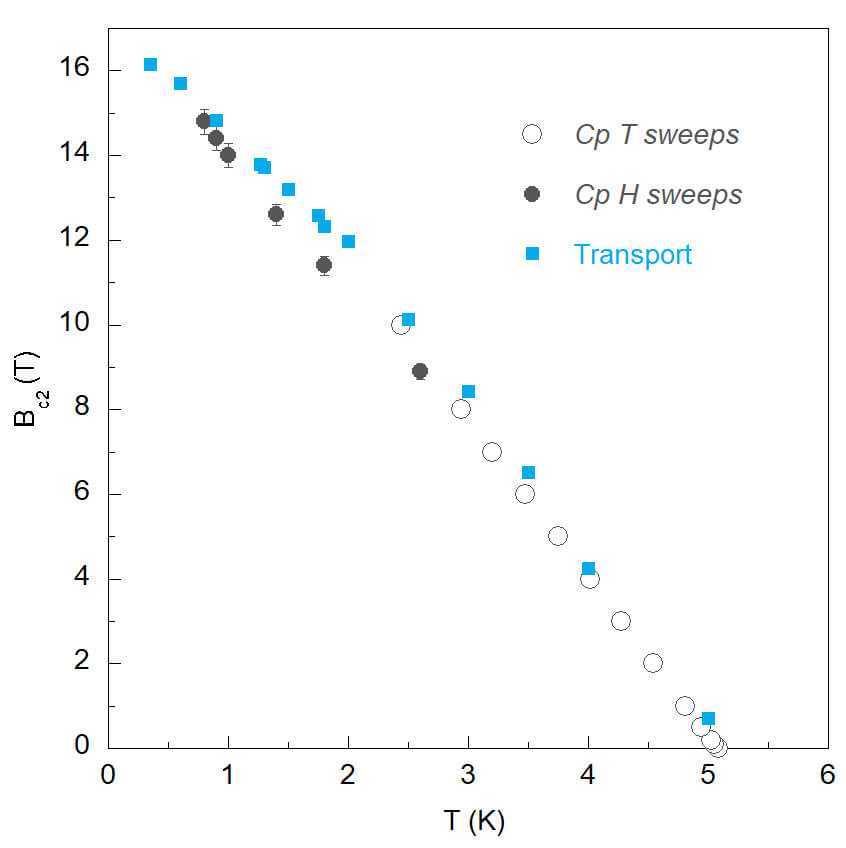}
\caption{Upper critical magnetic field vs temperature, as determined from
transport measurements (squares), as well as from heat-capacity measurements upon sweeping the temperature (mid-point of the transition - open circles), and upon sweeping the magnetic field (mid-point - full circles). }
\label{fig:Hc2}
\end{figure}

Resistivity shows a nearly temperature-independent behavior on cooling
from $T = 300$\,K to ca. 250~K. Below $T = 250$\,K, $\rho(T)$ shows an
activation-like behavior, as also reported previously~\cite{Gornicka2021}.
A possible explanation for this unexpected behavior is the formation of a
thin semiconducting/semi\-me\-tal\-lic oxide layer on an exposed sample surface.
This is consistent with the results of scanning tunneling spectroscopy measurements, which suggest that the immediate surface layer of the sample shows a lower $T_c$ and smaller energy gap than the bulk.
Nevertheless, the transition at the superconducting critical temperature $T_c$ = 5.2~K is relatively sharp, suggesting a good sample quality. The upper critical field $B_{c2}$ estimated from the resistivity measurements, here taken as the field corresponding to a 10$\%$ drop of the resistivity from its normal-state value, is plotted against temperature in Fig.~\ref{fig:Hc2} (blue squares).
Our data overlap with those from a previous report~\cite{Gornicka2021},
which where limited to 10\,T. The slope of $B_{c2}(T)$ was used to find
$B_{c2}(T = 0$\,K). Since this corresponds to 17.4\,T, this is 50\% above
the Pauli limit ($B_p = 11.6$\,T, if we take into account strong coupling).

STS measurements, realized at different fixed values of temperature and
magnetic field, allowed us to study the superconducting energy gap of
TaIr$_2$B$_2$. Prior to performing the spectroscopic measurements, we
tried to find areas of the sample that had a well-defined surface structure.
Our topographic measurements showed that due to ex-situ preparation,
the sample surface is strongly degraded and an overlying (insulating)
layer makes it impossible to study the topography of the surfaces.
Nevertheless, we were able to form tunnel junctions through this thin layer and locally measure the tunneling conductivity, which is proportional to the superconducting DOS.
All measured spectra show a single-gap character with two symmetric
gap-like peaks (see Fig.~\ref{fig:STM1}a). In some spectra other weak
features at higher voltages are observed. This is exemplified in the
middle spectrum of Fig.~\ref{fig:STM1}a, where a weak maximum is seen
at about 1\,mV. Such situation recalls some previous studies of MgB$_2$,
the iconic two-gap superconductor. In that case, the first STM spectra,
measured on MgB$_2$ powder~\cite{Bollinger2001} exhibited a small
superconducting energy gap, yet a weak maximum at higher energies went
unnoticed. Only later such a maximum was assigned to a second, large
energy gap by our group~\cite{prl2001} and others. The apparently single-gap spectrum
of MgB$_2$ occurs because the large gap on the two dimensional $\sigma$-band
is observed only when there is a tunneling component in the $ab$-planes,
while the small gap on the $\pi$-band is always present. However, this
is not the case here. On the polycrystalline TaIr$_2$B$_2$ samples we measured
hundreds of spectra on randomly oriented crystalline grains, but could
only see the manifestation of a single energy gap of about 0.25--0.45\,meV,
while rarely observed other, much weaker maxima occur at random positions.
If the gaps presented in Fig.~\ref{fig:STM1}a were related to the small
energy gap of a two-gap spectrum, this must show up in measurements in
a magnetic field. Indeed, in the MgB$_2$ case, the small gap is suppressed
at a much smaller critical field than the large one. In Ref.~\cite{Samuely2003}
it was shown that, on the seemingly single-gap spectrum with a prevailing
$\pi$-band gap, a rapid suppression of the small gap in a magnetic field
was accompanied by a shift of the gap-like coherent peaks toward higher
values of the large gap. Albeit small in intensity, the latter gap persists
at much higher magnetic fields. However, no such effect is observed in
TaIr$_2$B$_2$ (see also Fig.~S4).

Comparison of the data with the BCS theory revealed a good agreement with the single-gap $s$-wave model (see red dashed lines in Fig.~\ref{fig:STM1}a). However, the values of the energy gap $\Delta (0)$ derived from the fitting procedure vary between 0.25\,meV and 0.46\,meV, which are significantly lower than what is expected for a superconductor with a critical temperature $T_c \sim 5$\,K, whose $\Delta (0)$ should be larger than 0.75\,meV. By measuring the tunneling spectra at various temperatures, we determine the local $T_c$. Figure~\ref{fig:STM1}b shows such a temperature dependence of the spectrum with $\Delta(0)=0.46$\,meV. We see that the gap-like peaks are suppressed at temperatures $T=3$--3.5\,K, but a reduced tunneling conductance indicates that the superconducting DOS survives up to bulk $T_c \sim 5$\,K. The temperature dependence of the energy gap $\Delta (T)$ was calculated from fitting the tunneling spectra at different temperatures. The resulting $\Delta (T)$ is shown in Fig.~\ref{fig:STM1}c by open symbols. Also the BCS-like prediction for $T_c = 5.1$\,K and a reduced gap is displayed by a
black dotted line. The experimental data do not follow it, but go significantly below the prediction.
The red dashed line shows the BCS dependence of $\Delta (T)$ with the coupling strength 2$\Delta/k_\mathrm{B}T_c = 4.4$, which was determined from the heat capacity measurements (see below). The experimental points follow this prediction up to 2\,K and then are directed towards bulk $T_c = 5.1$\,K. Such a behavior can be interpreted as a proximity effect between a degraded superconductor on the surface [with $\Delta (0) = 0.46$\,meV and $T_c = 2.5$\,K] and the bulk.
When approaching the surface with a lower $T_c$ the coherence length
becomes longer and the junction probes superconducting parts deeper in
the sample with higher $T_c$, eventually the bulk $T_c$. 
This is also evident from the magnetic field dependence of the tunneling spectra with $\Delta (0) = 0.26$\,meV at $T = 0.5$\,K, which are displayed in Fig.~S4 of the Supplemental Material~\cite{SM}. This local phase, which in the BCS limit should have a $T_c$ below 1.7\,K, exhibits pronounced superconducting features even at a magnetic field of 8\,T. As this exceeds by far the Pauli critical field of the phase with a small $T_c$, it indicates the presence of a proximity effect from the bulk phase. Overall, STM suggests that tunneling spectra correspond to the surface proximity layer, where the superconductivity is induced from the $s$-wave single-gap bulk phase.

\begin{figure*}[th]
\centering
\vspace{-2mm}
\raisebox{2mm}{\includegraphics[width=0.423\textwidth]{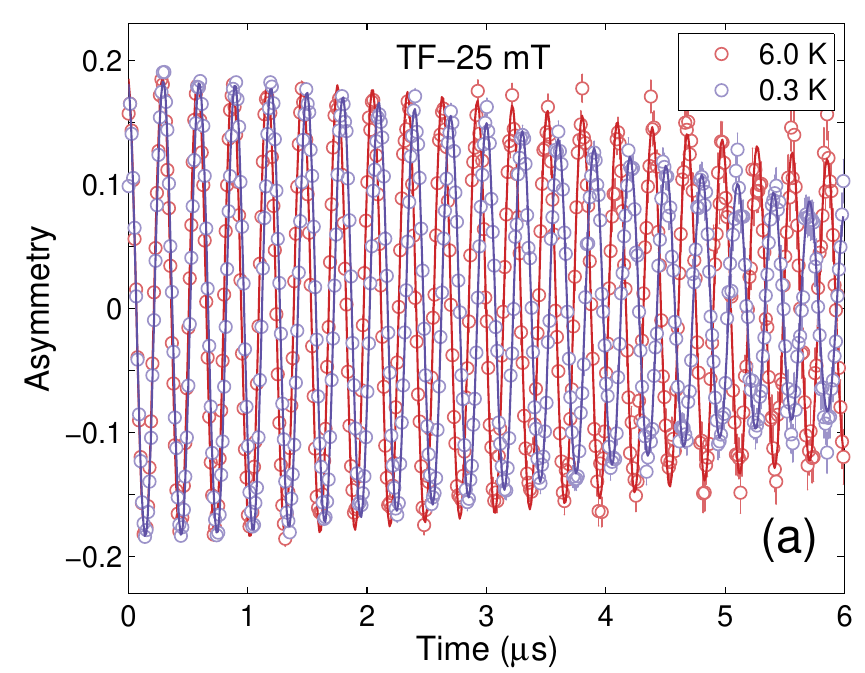}}
\hspace{3ex}
\includegraphics[width=0.44\textwidth]{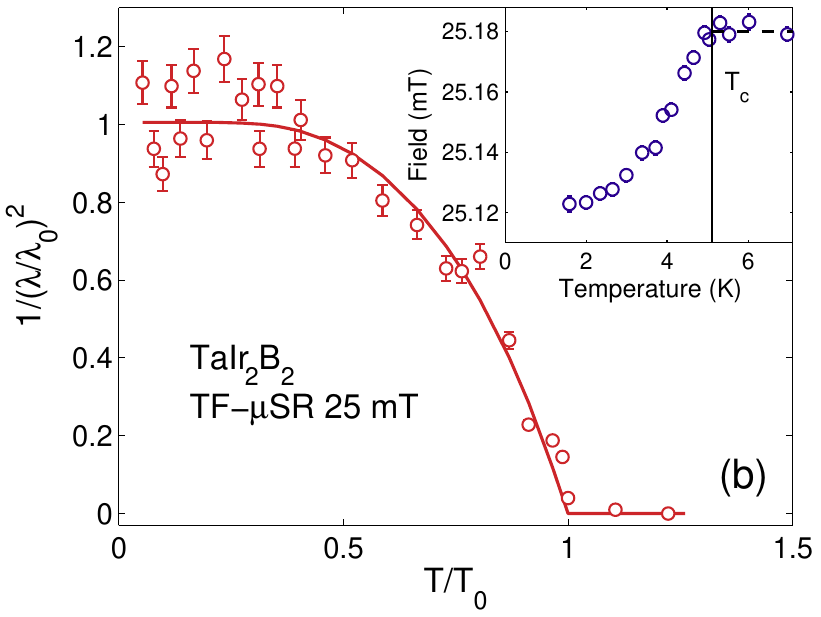}
\caption{\label{fig:MuSR_spectra}a) TF-{\textmu}SR spectra collected in the superconducting- (0.3\,K) and the normal state (6\,K) of TaIr$_2$B$_2$ in an applied magnetic field of 25\,mT. Solid lines represent fits to Eq.~(\ref{eq:asym}). At 0.3\,K, the increased decay and the reduced frequency reflect the onset of the vortex lattice and the diamagnetic shift in the superconducting state. b) Superfluid density vs temperature, as determined from TF-{\textmu}SR measurements at 25\,mT. The line represents a fit to a fully-gapped $s$-wave model with a single superconducting energy gap with $2\Delta/k_\mathrm{B}T_c = 4.5$. Inset: temperature dependence of the diamagnetic shift in the SC state.}
\end{figure*}

We also performed {\textmu}SR experiments. Representative time-domain spectra, recorded in the normal- and superconducting state of TaIr$_2$B$_2$ in an applied field of 25 mT, are shown in Fig.~\ref{fig:MuSR_spectra}a. The decay of asymmetry in the normal state is due to the muon-spin depolarization by the nuclear moments. Since such contribution is mostly constant in the considered temperature range, any additional increase in the {\textmu}SR decay rate reflects the field modulation due to the superconducting vortices. In particular, the {\textmu}SR asymmetry damping in the superconducting vortex state is determined by the width of the field distribution, in turn depending on the screening length scale $\lambda$. At the same time, the average muon-spin precession frequency is directly proportional to the local magnetic field at the muon stopping site, which reflects the diamagnetic field shift associated with the supercurrents.
In the superconducting state, the generally asymmetric field distribution is well represented by a sum of several ($n$) Gaussians~\cite{Weber1993,Maisuradze2009a,Maisuradze2009b}: 
\begin{align}
    A(t) &= A_0 \sum_{i=1}^{n} r_i\,\exp\left(\frac{-\sigma_i^2 t^2}{2}\right) \cos{(\gamma_{\mu}{B_i}t + \phi)}, \label{eq:asym} \\ 
    \left<B\right> &= \sum_{i=1}^{n} r_i B_i \;{\mathrm{and}}\;
    \left<\Delta B^2\right> = \frac{\sigma^2}{\gamma_{\mu}^2} = \sum_{i=1}^{n} r_i \left[\frac{\sigma_i^2}{\gamma_{\mu}^2} + (B_i - \left<B\right>)^2 \right] \label{eq:deltaB}.
\end{align}
Here, $A_{0}$ is the initial asymmetry, and $\left< B \right>$ and $\left < \Delta B \right>$ describe the first- and second moment of the field distribution. In our case, it was sufficient to use $n$ = 2 below $T_{c}$ and $n$ = 1 above it.
The superconducting contribution to the second moment of the field distribution $\sigma_\mathrm{sc}$ = $\sqrt{\sigma^2-\sigma_n^2}$ was obtained  from the measured decay rate $\sigma$, by subtracting the contribution of the nuclear moments $\sigma_{n}$, with the latter being determined above $T_{c}$ and temperature independent. Within the approximation of a hexagonal Ginzburg-Landau vortex lattice, the relationship between $\sigma$ and the magnetic penetration depth $\lambda$ can be parameterized as~\cite{Brandt2003}:
\begin{equation}
   \sigma_\mathrm{sc}\,[\mu s^{-1}] = 4.854 \times 10^{4}(1-b)[1+1.21(1-\sqrt{b})^{3}] \lambda\,\mathrm{[nm]}^{-2},
\end{equation}
where $b = \left< B \right> /B_{c2}$ is the reduced magnetic field.

Fig.~\ref{fig:MuSR_spectra}b shows the temperature-dependent inverse square
of the magnetic penetration depth [proportional to the superfluid density,
i.e., $\lambda^{-2}(T) \propto \rho_\mathrm{sc}(T)$] for $\mu_{0}H = 25$\,mT. To analyze the $\rho_\mathrm{sc}(T)$ data we use a generic model described by:
\begin{equation}
	\label{eq:rhos}
	\rho_\mathrm{sc}(T) = 1 + 2\, \Bigg{\langle} \int^{\infty}_{\Delta_\mathrm{k}} \frac{E}{\sqrt{E^2-\Delta_\mathrm{k}^2}} \frac{\partial f}{\partial E} \mathrm{d}E \Bigg{\rangle}_\mathrm{FS}. 
\end{equation}
Here, $f = (1+e^{E/k_\mathrm{B}T})^{-1}$ is the Fermi function~\cite{Tinkham1996} and 
$\Delta_\mathrm{k}(T)$ 
is the superconducting 
gap function.
The temperature dependence of the gap is assumed to follow $\Delta(T) = \Delta_0 \mathrm{tanh} \{1.82[1.018(T_\mathrm{c}/T-1)]^{0.51} \}$~\cite{Tinkham1996,Carrington2003}, where $\Delta_0$ is the gap value at 0\,K.
As illustrated in Fig.~\ref{fig:MuSR_spectra}b, the temperature-invariant
superfluid density below $T_c$/3 strongly suggests the absence of
low-energy excitations and, hence, a fully-gapped superconducting state
in TaIr$_2$B$_2$. Consequently, the $\rho_\mathrm{sc}(T)$ is consistent
with an $s$-wave model, which describes the SC density data very well
across the entire temperature range and yields a zero-temperature gap
$\Delta_0$ = 2.25(4)\,$k_\mathrm{B} T_c = 0.98(7)$\,meV. 
Such SC gap value is consistent with the STM and heat-capacity results (see below) and, since $2 \Delta_0 / k_\mathrm{B} T_c = 4.5$ is clearly larger than the ideal BCS value (3.53), it suggests a strong electron-phonon coupling.
At the same time, the fit of the TF-{\textmu}SR 25-mT data provides $\lambda_0$ value of 920(10)\,nm, but this may be overestimated by few tens of percents,  reflecting that size of the sample grains is comparable to the penetration depth (see Fig.~S2 in Supplemental Material~\cite{SM}).
Note that the apparently large error bars at low temperature in Fig.~\ref{fig:MuSR_spectra}b are intrinsic and reflect the tiny change of depolarization from the normal- to the superconducting state (see Fig.~\ref{fig:MuSR_spectra}a).

\begin{figure}[th]
\centering
\includegraphics[width=0.4\textwidth]{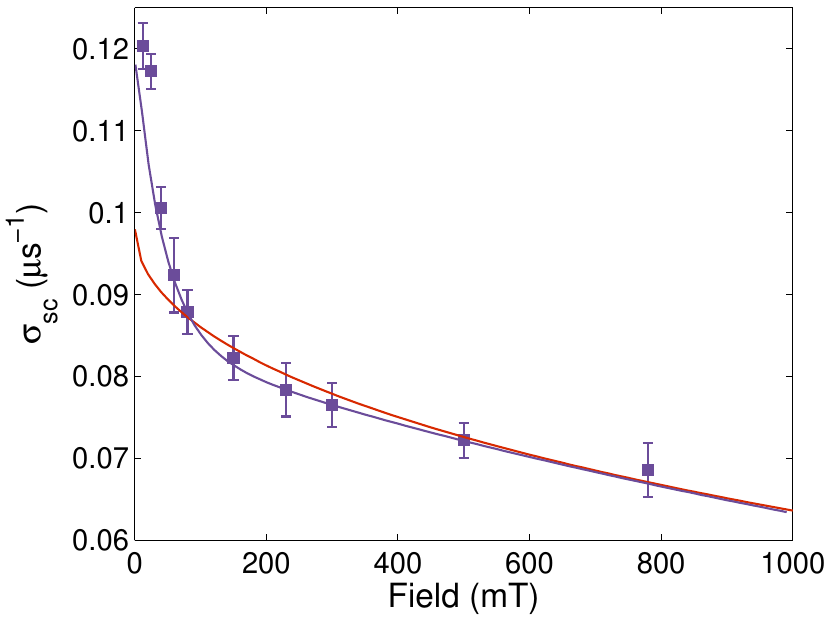}
\caption{\label{fig:sigma_vs_B}Field-dependent superconducting Gaussian relaxation rate $\sigma_\textrm{sc}(B)$ measured at 1.5\,K (on field cooling). The solid mauve line represents a fit to a two-band model with $w = 0.5$, $\xi_1 = 33.5$\,nm,  $\xi_2 = 4.4$\,nm, and $\lambda = 925$\,nm. Note the almost 8:1 ratio of coherence lengths $\xi$ in TaIr$_2$B$_2$. The red line shows a single-band model, which does not fit well the low-field data.}
\end{figure}

Since the $\sigma$-vs-$T$ dependence sometimes provides ambiguous outcomes regarding the single- vs multigap nature of superconductivity, we performed also field-dependent $\sigma$-vs-$B$ measurements in the superconducting state (at 1.5\,K), as shown in Fig.~\ref{fig:sigma_vs_B}. 
Remarkably the field dependence shows a strong change of slope at about 0.1\,T. Such a dependence cannot be fitted by a single-band formula (red line). On the contrary a perfect fit is achieved by a two-band model including two coherence lengths $\xi_1 = 33.48$\,nm and $\xi_2 = 4.4$\,nm, with weights $w$ and $(1-w)$~\cite{Serventi2004,Shang2022,Khasanov2014}. While $\xi_2$ refers to the bulk upper critical magnetic field $B_{c2} = 17$\,T, $\xi_1$ refers to a small field of 0.3\,T. All this suggest a presence of two-band effects in the system, see discussion below.

\begin{figure}[th]
\centering
\includegraphics[width=0.4\textwidth,angle=0]{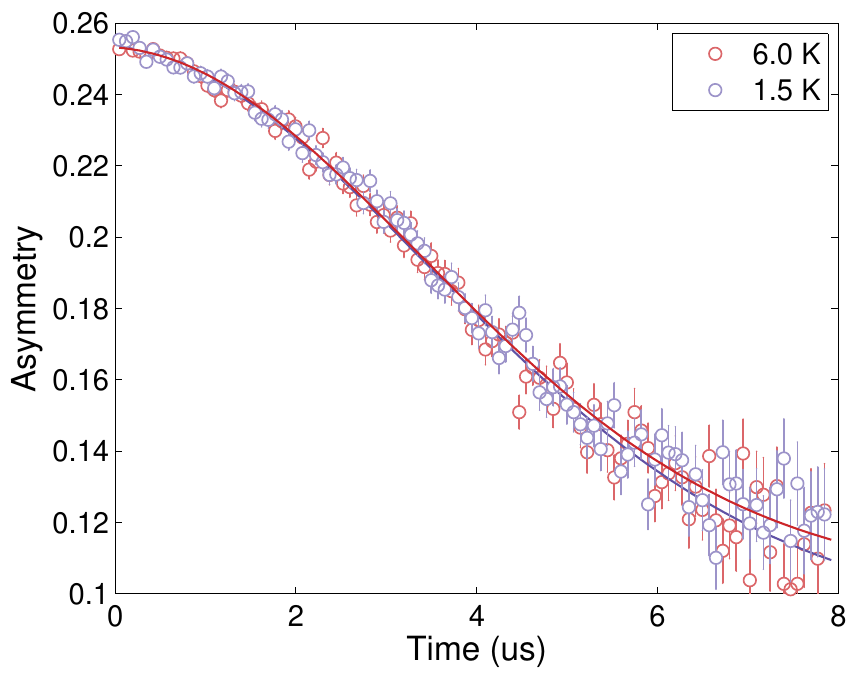}
\caption{\label{fig:ZF-muSR}ZF-{\textmu}SR spectra of a TaIr$_2$B$_2$ in the
superconducting- (1.5\,K) and in the normal state (6\,K).
The solid lines are fits with a combined Gaussian- $\sigma_\mathrm{ZF}(T)$
and Lorentzian $\Lambda_\mathrm{ZF}(T)$ relaxation function, accounting for the nuclear- and electronic contributions, respectively.
The almost overlapping fits and datasets suggest a preserved TRS.}
\end{figure}

To search for a possible breaking of the time-reversal symmetry in
TaIr$_2$B$_2$, we performed zero-field (ZF-) {\textmu}SR measurements in its
normal- and superconducting state. As shown in Fig.~\ref{fig:ZF-muSR},
in either case, neither coherent oscillations nor a fast decay could
be identified, thus implying the lack of any magnetic order or fluctuations. 
The muon-spin relaxation in TaIr$_2$B$_2$ is mainly due to the randomly
oriented nuclear moments (mostly $^{11}$B and $^{181}$Ta), which can be
modeled by a Gaussian Kubo-Toyabe relaxation function
$G_\mathrm{KT} = [\frac{1}{3} + \frac{2}{3}(1 -\sigma_\mathrm{ZF}^{2}t^{2})\,\mathrm{e}^{-\sigma_\mathrm{ZF}^{2}t^{2}/2}]$~\cite{Kubo1967,Yaouanc2011}, with $\sigma_\mathrm{ZF}$
the zero-field Gaussian relaxation rate. 
The ZF-{\textmu}SR spectra were fitted by considering also an additional
electronic contribution $\mathrm{e}^{-\Lambda_\mathrm{ZF} t}$.
The solid lines in Fig.~\ref{fig:ZF-muSR} represent fits to
$A_\mathrm{ZF}(t) = A_\mathrm{s} G_\mathrm{KT} \mathrm{e}^{-\Lambda_\mathrm{ZF} t} + A_\mathrm{bg}$, with $\Lambda_\mathrm{ZF}$ the zero-field exponential muon-spin relaxation rate, with $A_\mathrm{s}$ and $A_\mathrm{bg}$ being the same as in TF-{\textmu}SR. 
Since neither $\Lambda_\mathrm{ZF}(T)$ nor $\sigma_\mathrm{ZF}(T)$ show a systematic enhancement below $T_c$, this excludes a possible breaking of TRS in the superconducting state of
TaIr$_2$B$_2$. Such circumstance is reflected in the almost
overlapping datasets shown in Fig.~\ref{fig:ZF-muSR}.

\begin{figure}[th]
\centering
\includegraphics[width=0.36\textwidth]{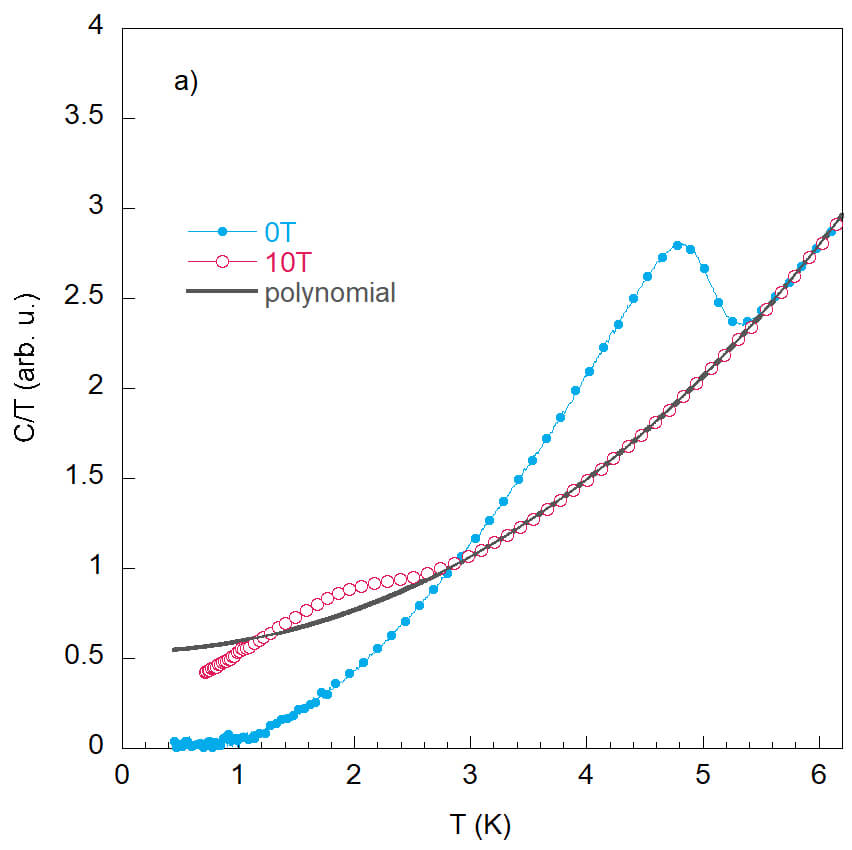}
\hspace{5mm}
\includegraphics[width=0.36\textwidth]{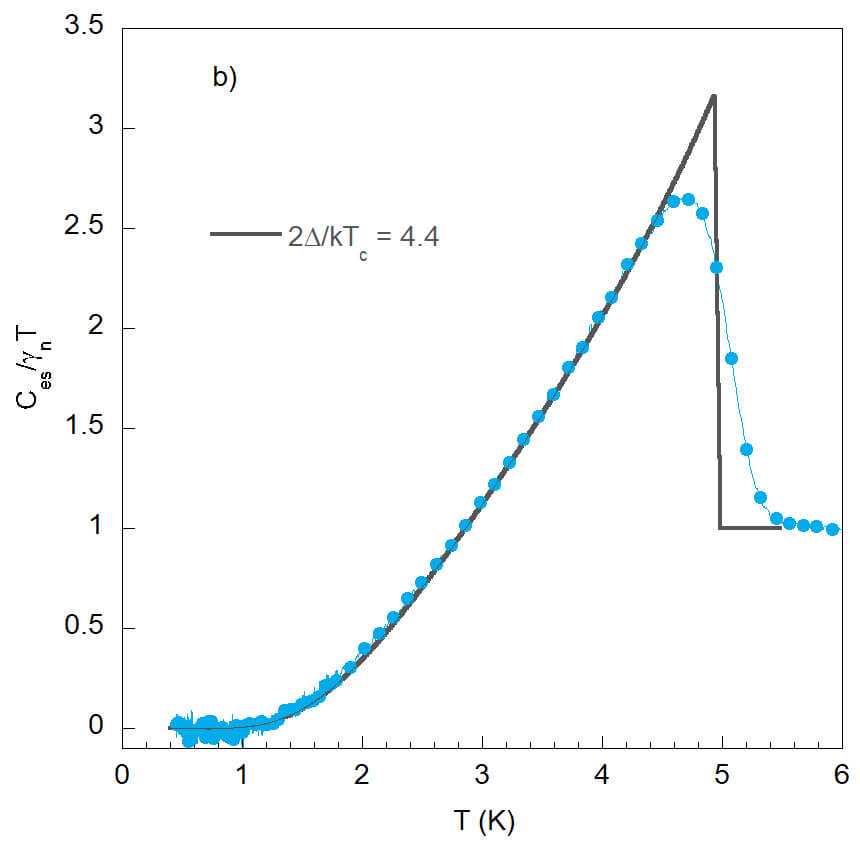}
\caption{a) Heat capacity of TaIr$_2$B$_2$ measured at 0\,T (line
+ full symbols) and 10\,T (line + empty symbols). For clarity, in both
cases only every 30th measured point is displayed as a symbol. The thick
line is a polynomial fit to the 10-T data above the superconducting
transition. b) Electronic heat capacity (line + symbols) and a fit
using the alpha-model (thick line).}
\label{fig:Cp1}
\end{figure}

In the following, we present the analysis of the heat-capacity data. Figure~\ref{fig:Cp1}a shows the total heat capacity of the sample at 0 and 10\,T. Note that,  for clarity, only one point out of 30 is displayed. At 0\,T, $C_{s}$(0\,T)/$T$ shows a relatively sharp superconducting transition close to 5\,K, followed by an exponential decrease with decreasing temperature, reproducing previous findings, reported only in a limited temperature range~\cite{Gornicka2021}. At 10\,T,
the superconducting transition is still visible, but shifted to lower temperatures. Obviously, a magnetic field of 10\,T is not sufficient to entirely suppress
the superconductivity. Therefore, the normal-state heat capacity $C_{n}/T$ needed to be calculated following a polynomial fit. For this, a standard formula was used, including electronic and phononic contributions in the form $C_{n} = aT + bT^3 + cT^5$. The fit result, extrapolated down to low temperatures, is depicted as a continuous line in Fig.~\ref{fig:Cp1}a.

\begin{figure*}[th]
\centering
\includegraphics[width=0.36\textwidth]{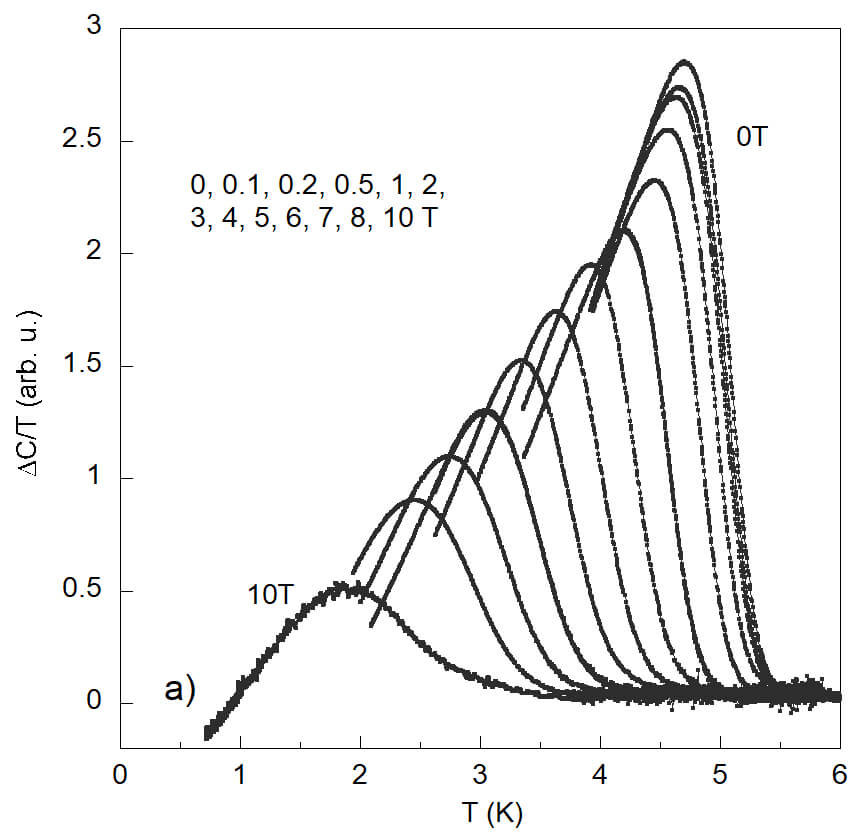}
\hspace{8mm}
\includegraphics[width=0.36\textwidth]{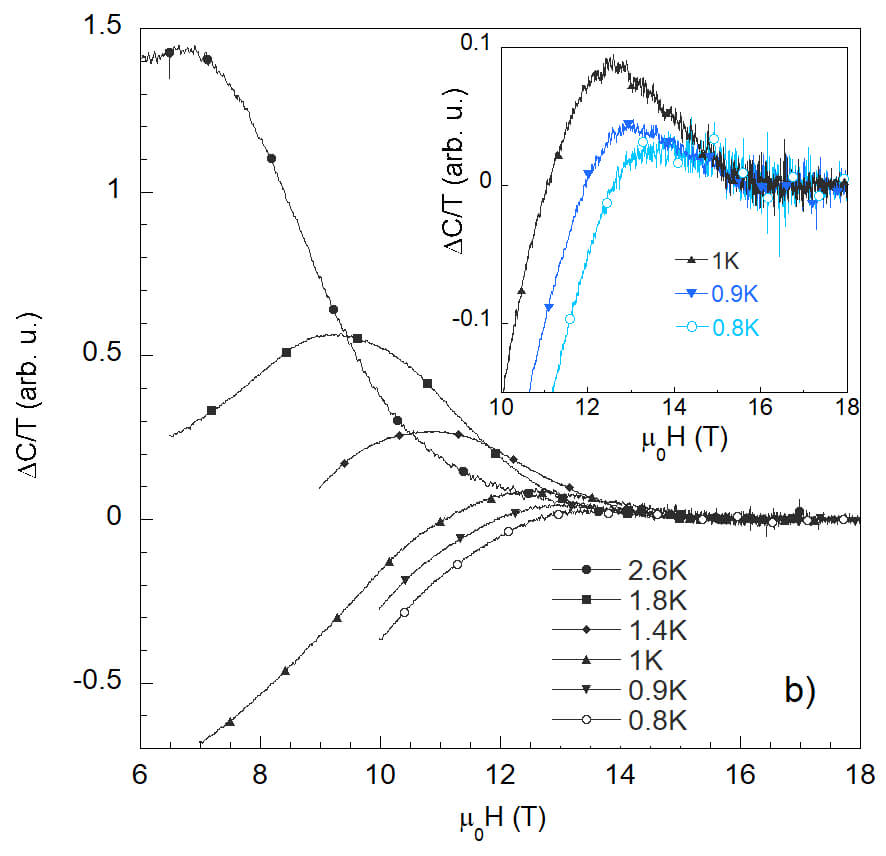}
\caption{a) Heat capacity measurements in various fixed magnetic fields after subtraction of the normal state contribution. b) Heat capacity measurements at various fixed temperatures after subtraction of the normal state contribution. Inset: detail of selected curves from the main panel.}
\label{fig:Cp2}
\end{figure*}

In order to analyze the heat capacity of TaIr$_2$B$_2$, we calculated the electronic heat capacity of the sample as $C_{es}/\gamma_nT=\Delta C/\gamma_nT$ + 1, where $\Delta C/T=(C_{s}(0\,\mathrm{T})-C_{n})/T$, and $\gamma_n=C_n/T_{0\,\mathrm{K}}$. The result is depicted in Fig.~\ref{fig:Cp1}b by blue points. The anomaly at the transition is well pronounced and rather sharp, proving the good quality of the sample. To estimate the coupling strength of the superconducting electrons we compared $C_{es}/\gamma_nT$ with the so-called alpha-model~\cite{Padamsee} based on the BCS theory. The only parameter in this model is the ratio of the superconducting energy gap $\Delta$ to the critical temperature $T_c$, $2\Delta/k_\mathrm{B}T_c$. If necessary, the model may be also adjusted to
account for two-gap superconductivity (see for example the case of
MgB$_2$~\cite{Fisher} or NbS$_2$~\cite{Kacmarcik}) or for an anisotropic
energy gap. Here, we find a very good agreement between the experimental
data and the single-gap model for a ratio $2\Delta/k_\mathrm{B}T_c = 4.4 \pm 0.2$,
pointing to a strong coupling in the system. In the figure, the fit with $2\Delta/k_\mathrm{B}T_c = 4.4$ is displayed as a thick line. It is clear that the single-gap model describes quite accurately the data over the entire temperature range, without the need to include a second energy gap. The entropy conservation is satisfied in the overall temperature range, proving the thermodynamic consistency of our measurements. From the fit, the bulk $T_c$ in a zero magnetic field is 4.95\,K.

We have obtained $\Delta C/T$ for a series of fixed magnetic-field measurements. The results are summarized in Fig.~\ref{fig:Cp2}a. The transition gradually shifts 
to lower temperatures and the jump at $T_c$ progressively diminishes as the field increases from 0 to 10\,T. Taking the critical temperatures for each curve, measured in a distinct magnetic field, it is possible to construct the temperature dependence of the upper critical field $B_{c2}$. For simplicity, we
derive $B_{c2}$ from the mid-point of each heat-capacity transition. The result is presented in Fig.~\ref{fig:Hc2} by full circles and reveals a linear increase with decreasing temperature in the considered magnetic field range. In the figure, the results of the upper critical magnetic field derived from the transport measurements are shown as well, displayed as filled blue squares. From the transport measurements, the linear increase of $B_{c2}$ appears to continue down to the lowest temperatures. To confirm this trend via heat-capacity measurements at high magnetic fields, we performed further $C$ measurements at EMFL Grenoble, France. To detect the superconducting transition at high magnetic fields it is more convenient to run the measurement at a fixed temperature and sweep the magnetic field. The obtained heat-capacity data, after the subtraction of the corresponding normal-state contribution, are depicted in Fig.~\ref{fig:Cp2}b. The inset of the figure shows in detail three low-temperature datasets.
From the magnetic field sweeps, the upper critical field was
extracted as the midpoint of each transition, as shown in Fig.~\ref{fig:Hc2}
with open circles. This confirms the linear temperature dependence down
to the lowest temperatures. However, a strict claim of linearity should
be taken with a grain of salt, as the transitions in the heat capacity
become very broad upon cooling. On the other hand, the linearity appears
to be robust regardless of the criterion used to define $B_{c2}$.
For comparison, we determined the position of the maximum at the
transition and report it with open crosses in Fig.~S5 of the Supplemental
Material~\cite{SM}. Additionally, the $B_{c2}$ values taken at 10\%, 50\%,
and 90\% of the resistive transition are also presented in Fig.~S5.
Thus, the high-field measurements confirm the nontrivial linear behavior
of the upper critical field and the high $B_{c2}$ value in TaIr$_2$B$_2$,
well above the Pauli limit.

\section{Discussion}
First, we address the issue of the order parameter of TaIr$_2$B$_2$. Both, the temperature-dependent heat capacity and superfluid density show consistently an $s$-wave single-gap behavior with a strong coupling ratio 2$\Delta/k_\mathrm{B}T_c \approx 4.4$. This is also supported by the STM spectroscopy data.  On the other hand, the field-dependent superconducting Gaussian relaxation rate $\sigma_\textrm{sc}(B)$ can be described only by two coherence lengths $\xi$, differing by almost a factor of 8 and with comparable weights. The small $\xi$ relates to the bulk upper critical field $B_{c2} = \Phi_0/2\pi \xi ^2 = 17$\,T while the large one to $B_{c2} = 0.3$\,T. This reminds the case of MgB$_2$, where the superconducting coupling on the $\pi$-band is much more sensitive to applied magnetic field and the small gap $\Delta_{\pi}$ closes at a field much lower than the bulk $B_{c2}$~\cite{prl2001}. 
Hence, in TaIr$_2$B$_2$, the existence of a small energy gap should be considered. The coherence length is given as  $\xi\sim v_\mathrm{F}/\Delta (0)$, with $v_\mathrm{F}$ the Fermi velocity. Let us first assume that the $v_\mathrm{F}$ values  corresponding to the different bands are similar. Then, the two energy gaps should differ in the same way as the large- and small $\xi$, i.e., by a factor of 8. To check if such a combination of energy gaps and weights 
is beyond our experimental resolution, so as to leave the smaller energy gap unnoticed, we calculated the two-gap model curves for the temperature dependence of the heat capacity and superfluid density, as well as the two-gap STM superconducting spectrum. The resulting model curves are displayed in the Supplemental Material~\cite{SM} as Figs.~S6, S7, and S8, respectively. As can be seen in the heat-capacity data, the contribution of such a small energy gap would become important only well below 1\,K, where our experimental uncertainty increases significantly and the small gap cannot be completely excluded. On the other hand, based on a dirty-limit model for the superfluid density, a linear combination of two contributions, corresponding to a small- and a large energy gap with similar weights, leads to a pronounced feature below 1\,K (Fig.~S7 in \cite{SM}), which is clearly absent in the experimental data shown in Fig.~\ref{fig:MuSR_spectra}b. Also a linear combination of two superconducting density of states in the tunneling conductance, with one large gap and another gap by 1/8 smaller, clearly reflects the small gap, which again is not observed in the experimental data displayed in Fig.~\ref{fig:STM1}a. Thus, most likely, TaIr$_2$B$_2$ is an $s$-wave single-gap superconductor with a strong superconducting coupling. In the expression  $\xi\sim v_\mathrm{F}/\Delta (0)$, different Fermi velocities on different bands may be considered to resolve the apparent contradiction between the direct gap measurements and the strongly field dependent muon-spin relaxation rate $\sigma_\mathrm{sc}$. Then, TaIr$_2$B$_2$ seems to have two bands with distinct coherence lengths, but the related Fermi-surface sheets have equal superconducting gaps.
A very similar situation appears to occur in SrPt$_3$P~\cite{Khasanov2014},
where $\mu$SR experiments suggest the presence of equal gaps but with different coherence lengths across the two Fermi-surface sheets. $\beta$-Bi$_2$Pd
is another example, where STM data indicate a single isotropic energy gap
as well as significant multiband effects in the mixed-state properties,
such as an enhanced intervortex tunneling conductance and an unusual behavior
of the upper critical magnetic field~\cite{Herrera2015}.

In the following we discuss the two issues related to the upper critical field of TaIr$_2$B$_2$ --- the linear behavior of $B_{c2} (T)$ and the high $B_{c2}(0)$ value, well above the Pauli limit. If we assume the strong coupling of $2\Delta/k_\mathrm{B}T_c = 4.4$, found by our specific-heat and {\textmu}SR measurements, and a $g$-factor $g = 2$, the Pauli limiting field is $B_\mathrm{P} = 11.9$\,T. From Fig.~\ref{fig:Hc2}, the extrapolated zero-temperature $B_{c2} (0) = 17.4$\,T is 50\% stronger, clearly indicating that the standard paramagnetic pair-breaking effect is suppressed. This may be related to the presence of the Ta and Ir heavy atoms, with atomic numbers $Z = 73$ and 77, respectively. Spin-orbit scattering has been shown to follow the Abrikosov–Gor’kov value $\hslash/\tau_\mathrm{SO} \propto Z^4$, which can lead to a short spin-orbit scattering time $\tau_\mathrm{SO}$, making the spin not a good quantum number. Similar to the TaIr$_2$B$_2$ case, Coleman et al.~\cite{Coleman1983} in intercalated TaS$_2$ and Sacépé et al.~\cite{sacepe2015}  in strongly disordered InO$_x$ have shown that a short $\tau_\mathrm{SO}$ leads to a suppression of the  paramagnetic pair-breaking, thus allowing for very high $B_{c2}(0)$ values.

The majority of classical type-II superconductors is characterized by a temperature dependence of the upper critical field that follows the standard semi-classical Werthamer-Hel\-fand-Hohenberg (WHH) theory~\cite{WHH,WHH2}. Such theory predicts the saturation of $B_{c2}$ at low temperature, differing slightly depending on spin paramagnetic and spin-orbit effects, respectively. Yet, several classes of superconductors exhibit an anomalous enhancement of $B_{c2} (T) $ at low temperatures and, specifically, a linear temperature dependence of $B_{c2}$
down to very low temperatures. A linear $B_{c2}$($T$) occurs in some multi-gap superconductors such as pnictides~\cite{Zhang2011,Hanisch}. As treated theoretically, e.g., by Gurevich~\cite{Gurevich} it is due to interplay of different gaps on different bands and the intraband and interband scatterings. 
Strong coupling effects combined with disorder may be also at play.
Kuchinskii et al.~\cite{Kuchinskii} have shown that, in an attractive Hubbard model, in the region of BCS- to the Bose-Einstein crossover, i.e., for a strong coupling and strong disorder, the $B_{c2}$($T$) dependence becomes practically linear.
Heavily disordered amorphous transition-metal alloys, as Mo$_{30}$Re$_{70}$, with d$B_{c2}$/d$T = -3.6$\,T/K~\cite{Tenhover1981}, exhibit an increased linearity upon increasing the normal-state resistivity close to the electron localization.
Recently, the quasicrystal superconductor Ta$_{1.6}$Te has been shown to
have an upper critical field which increases linearly with a slope of $-4.4$\,T/K with decreasing temperature (down to 40\,mK), with no tendency to level off. Its extrapolated zero-temperature critical field exceeds the Pauli limit by a factor of 2.3~\cite{Terashima}. A nonuniform superconducting gap and spin-orbit interactions in quasicrystal structures were considered. Sac\'ep\'e et al.~\cite{Sacepe2019} found that, in disordered amorphous indium oxide and in MoGe films at low temperatures, $B_{c2}$($T$) deviates from saturation towards linearity, a tendency which is reinforced upon increasing the sheet resistance between 0.5 and 3.5\,k$\mathrm{\Omega}$. Their theoretical explanation based on a vortex-glass ground state and its thermal fluctuations could be applicable to both films and disordered bulk superconductors with a low superfluid density, where the phase fluctuations are strong.

Based on evidence gained in this study, it can be stated that TaIr$_2$B$_2$ is not a multi-gap superconductor, but it still has a multi-band character. Electronic band-structure calculations~\cite{Gornicka2021} show that, due to spin-orbit coupling and a lack of inversion symmetry, the two bands crossing the Fermi level split. In this case, the energy-band splitting $E_\mathrm{ASOC}$ depends strongly on momentum $k$ and it ranges from 25 to 250\,meV, as large as in CePt$_3$Si~\cite{Smidman2017}. SOC not only splits each Fermi surface into two sheets but, most importantly, it changes their topology, so that bands with very different Fermi velocities appear. This can cause  strong multi(two)-band effects observed in the field-dependent superconducting muon relaxation rate $\sigma_\textrm{sc}(B)$ and may also affect the unconventional linear temperature dependence of the upper critical magnetic field. Yet, strong coupling effects and disorder are also at play and can explain the linear $B_{c2}(T)$ behavior. Indeed, a high resistivity and its activation character suggest a high degree of disorder in TaIr$_2$B$_2$~\cite{Gornicka2021}. 
Comparative studies on single crystals with reduced disorder could shed light regarding which of the above effects is dominant. A theoretical treatment
of non-centrosymmetric superconductors with $B_{c2}(T)$ exceeding the Pauli limit is highly desirable as well.

\section*{Conclusions}
A comprehensive study of the non-centrosymmetric superconductor TaIr$_2$B$_2$ is presented employing transport, heat capacity, STM/STS and {\textmu}SR measurements. Heat-capacity and {\textmu}SR results consistently point to a single-gap superconductivity with a strong coupling ratio $2\Delta/k_\mathrm{B}T_c = 4.4$. The single energy gap is reflected also in the surface-sensitive scanning tunneling spectroscopy data, even if the resulting value of the energy gap is reduced compared to the bulk value. This is most probably related to the degradation of the sample surface, i.e., to the formation of an oxide layer on the exposed surface, where superconductivity is only induced from the bulk beneath via a proximity effect.
The strong field-dependence of the superconducting muon relaxation rate $\sigma_\textrm{sc}(B)$ is explained by the two-band effect, implying that TaIr$_2$B$_2$
is an example of a single-gap two-band superconductor.  

The upper critical magnetic field of TaIr$_2$B$_2$, determined from resistivity and heat capacity measurements down to very low temperatures, exceeds the Pauli limiting field by 50\% and increases linearly with decreasing temperature. This linear increase is most likely related to multiband effects, strong superconducting coupling and disorder.

\section*{CRediT authorship contribution statement}
\textbf{J.~Ka\v{c}mar\v{c}\'ik:} Data curation, Formal analysis.
\textbf{Z.~Pribulov\'a:} Data curation, Formal analysis, Writing – original draft, review \& editing
\textbf{T.~Shiroka:} Data curation, Formal analysis, Writing – original draft, review \& editing.
\textbf{F.~Ko\v{s}uth:} Data curation, Formal analysis.
\textbf{P.~Szab\'o:} Data curation, Formal analysis, Writing - original draft.
\textbf{M.~J.~Winiarski:} Data curation, Formal analysis.
\textbf{S.~Kr\'olak:} Data curation, Formal analysis.
\textbf{J.~Jaroszynski:} Data curation, Formal analysis.
\textbf{T.~Shang:} Data curation, Formal analysis.
\textbf{R.~J.~Cava:} Conceptualization.
\textbf{C.~Marcenat:} Data curation, Formal analysis.
\textbf{T.~Klein:} Formal analysis.
\textbf{T.~Klimczuk:} Conceptualization, Data curation, Formal analysis, Writing – original draft, review \& editing. 
\textbf{P.~Samuely:} Conceptualization, Formal analysis, Writing – original draft, review \& editing.

\section*{Declaration of competing interest}
The authors declare that they have no known competing financial
interests or personal relationships that could have appeared to
influence the work reported in this paper.

\section*{Acknowledgements}
The work at Gdansk University of Technology was supported by the National Science Center (Poland), Grant No.\ 2022/\-45/\-B/\-ST5/\-03916. The work at Slovak Academy of Sciences was supported by Projects APVV-23–0624, VEGA 2/0073/24, COST Action No.~CA21144 (SUPERQUMAP), Slovak Academy of Sciences Project IMPULZ IM-2021–42. We acknowledge the support of LNCMI-CNRS, member the European Magnetic Field Laboratory (EMFL). A portion of this work was performed at the National High Magnetic Field Laboratory, which is supported by National Science Foundation Cooperative Agreement No.~DMR-1644779 and the State of Florida. TS thanks R.\ Khasanov (PSI) for insightful discussions regarding the field-dependent {\textmu}SR data and acknowledges the beam time awarded at the S{\textmu}S facility (Dolly and GPS spectrometers).

\appendix
\section{\label{app:A}Supplementary data}
Supplementary data to this article can be found online at
https://doi.org/10.1016/j.supcon.2025.xxxxxx.

{\footnotesize 
\bibliography{ms_TaIr2B2}
}

\end{document}


\begin{frontmatter}

\makeatletter\renewcommand{\ps@plain}{%
\def\@evenhead{\hfill\itshape\rightmark}%
\def\@oddhead{\itshape\leftmark\hfill}%
\renewcommand{\@evenfoot}{\hfill\small{--~\thepage~--}\hfill}%
\renewcommand{\@oddfoot}{\hfill\small{--~\thepage~--}\hfill}%
}\makeatother\pagestyle{plain}


\newcommand{\pbcomment}[1]{{\color{red} #1}}

\title{Supplementary material for\\
Single-gap two-band superconductivity well above the Pauli limit in non-centrosymmetric TaIr\textsubscript{2}B\textsubscript{2}}
\author[1]{J.~Ka\v{c}mar\v{c}\'ik}
\author[1]{Z.~Pribulov\'a}
\author[2,3]{T.~Shiroka\corref{cor1}}
\ead{tshiroka@phys.ethz.ch}
\author[1,4]{F.~Ko\v{s}uth}
\author[1]{P.~Szab\'o}
\author[5]{M.~J.~Winiarski}
\author[5]{S.~Kr\'olak}
\author[6]{J.~Jaroszynski}
\author[7]{T.~Shang}
\author[8]{R.~J.~Cava}
\author[9]{C.~Marcenat}
\author[10]{T.~Klein}
\author[5]{T.~Klimczuk\corref{cor2}}
\ead{tomasz.klimczuk@pg.edu.pl}
\author[1]{P.~Samuely}
\address[1]{Centre of Low Temperature Physics, Institute of Experimental Physics, Slovak Academy of Sciences, SK‑04001 Košice, Slovakia}
\address[2]{Laboratorium f\"{u}r Festk\"{o}rperphysik, ETH Zürich, 8093 Zurich, Switzerland}
\address[3]{PSI Center for Neutron and Muon Sciences CNM, 5232 Villigen PSI,  Switzerland }
\address[4]{Centre of Low Temperature Physics, Faculty of Science, P. J. Šafárik University, SK-04001 Košice, Slovakia}
\address[5]{Faculty of Applied Physics and Mathematics and Advanced Material Center, Gdansk University of Technology, ul. Narutowicza 11/12, Gdańsk 80–233, Poland}
\address[6]{National High Magnetic Field Laboratory, Florida State University, Tallahassee, Florida 32310, USA}
\address[7]{Key Laboratory of Polar Materials and Devices (MOE), School of Physics and Electronic Science, East China Normal University, Shanghai 200241, China}
\address[8]{Department of Chemistry, Princeton University, Princeton, 08544, New Jersey, USA}
\address[9]{Univ. Grenoble Alpes, CEA, Grenoble INP, IRIG, PHELIQS, 38000, Grenoble, France}
\address[10]{Univ. Grenoble Alpes, CNRS, Institut Néel, 38000, Grenoble, France}

\cortext[cor1]{Corresponding author}
\cortext[cor2]{Corresponding author}

\date{\today}
\vspace{-3mm}
\end{frontmatter}


\section*{Structural analysis}
The pXRD data were analyzed by the Rietveld method with the starting model described in Ref.~[1].
The obtained lattice parameters are: $a = 8.1318(4)$\,\AA, $b = 4.76430(7)$\,\AA, $c = 6.0082(2)$\,\AA, $\beta = 102.146(1)^{\circ}$, in very good agreement with the data reported in Ref.~[1].

\begin{figure}[ht]
\centering
\includegraphics[width=0.40\textwidth]{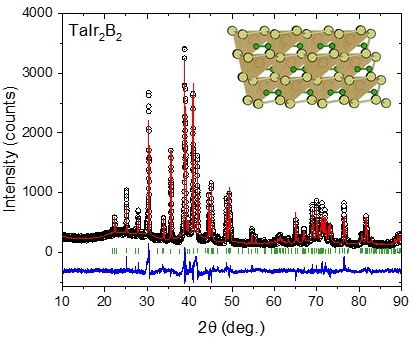} 
\vspace{-3mm}
\caption{Powder X-ray diffraction pattern for TaIr$_2$B$_2$ (black points),
refined using the Rietveld method (red line) with $R_w = 11.19$ and
$\chi^2 = 3.19$. The model assumes the space group $Cc$ and the atomic positions shown in Table~1. The inset presents the crystal structure of TaIr$_2$B$_2$, with the Ta-centered polyhedra (brown) surrounded by Ir atoms (yellow), and the boron dimers (green) filling the voids.}
\label{fig:XRD}
\end{figure}

\fontsize{8}{9}\selectfont
\noindent
[1] K.~Górnicka et al.,
Adv. Funct. Mater. 31 (2021) 2007960.
\fontsize{10}{12}\selectfont{}

\begin{figure}[ht]
\centering
\includegraphics[width=0.27\textwidth]{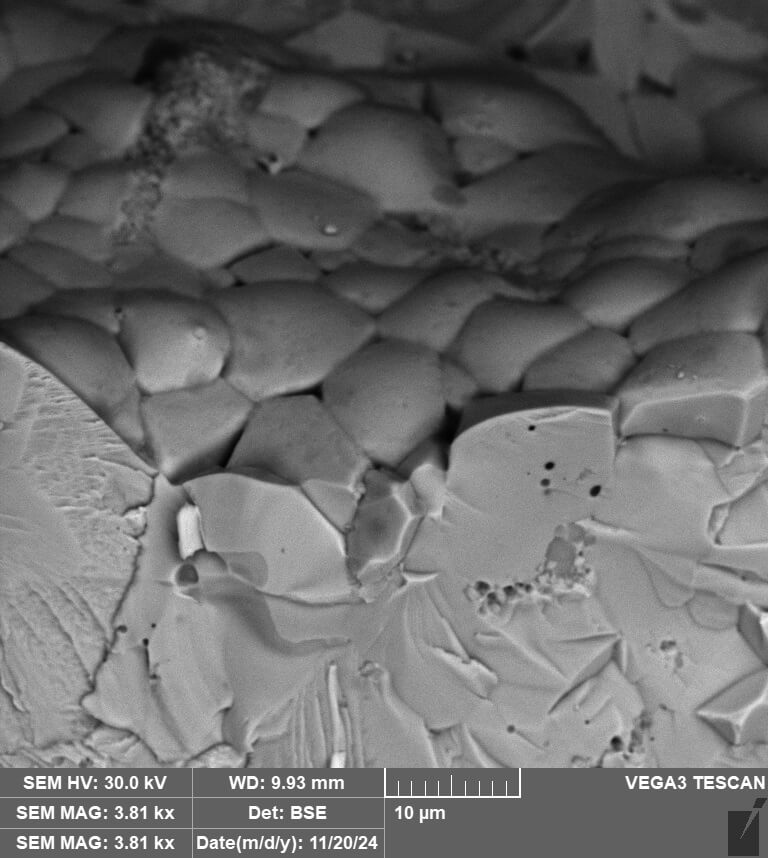}
\caption{Scanning eletron microscopy analysis of the sample surface reveal a typical grain size of a few micrometers.}
\label{fig:SEM}
\end{figure}

\section*{Transport}
\begin{figure}[hb!]
\centering
\includegraphics[width=0.35\textwidth]{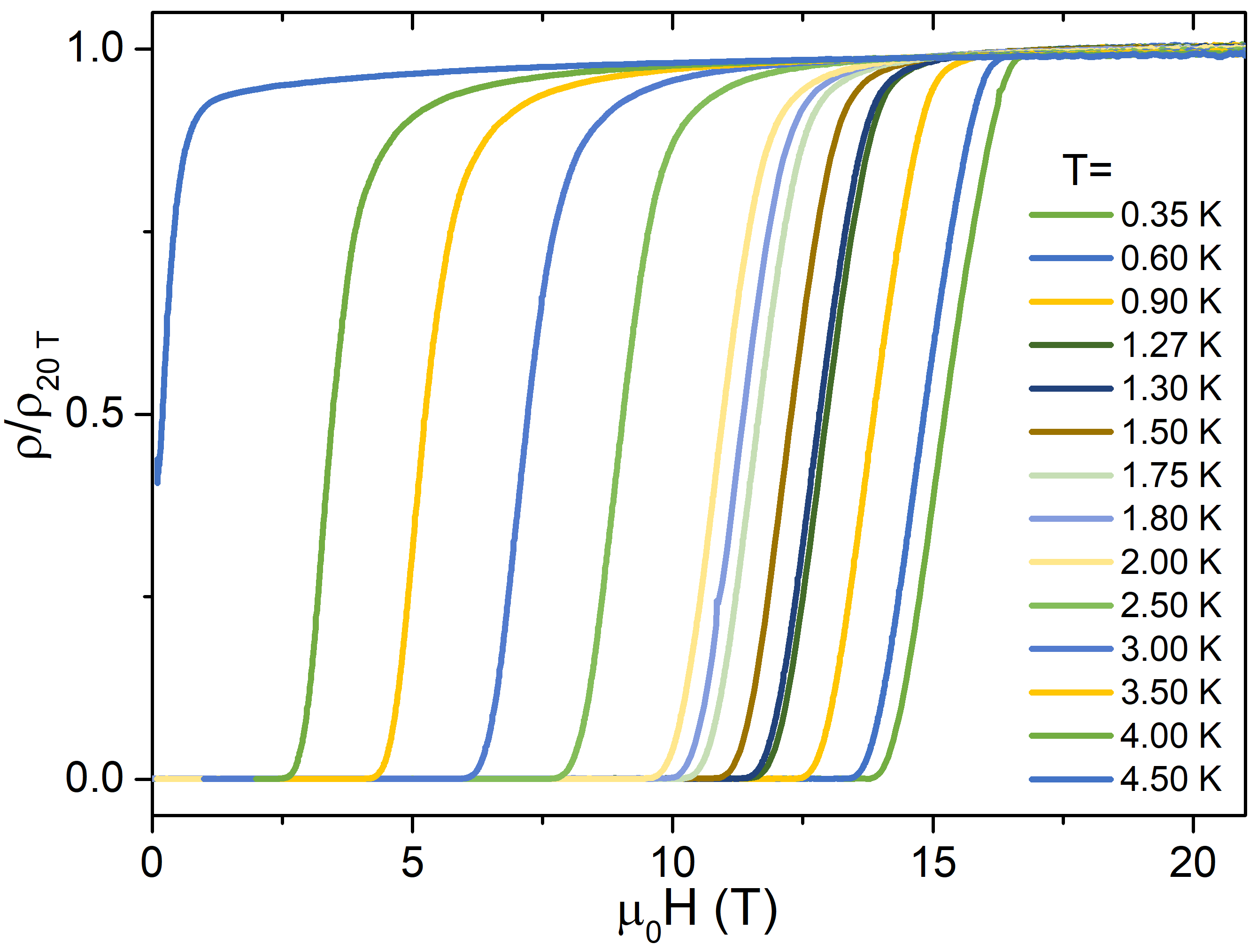}
\vspace{-3mm}
\caption{The electrical resistivity of TaIr$_2$B$_2$ as a function of
magnetic field measured at various fixed temperatures.}
\label{fig:RvsH}
\end{figure}
\vfill

\newpage

\section*{STM}
STM measurements show consistently single-gap spectra, but due to surface degradation local critical temperature, as well as the energy gap is reduced compared to bulk.
\begin{figure}[ht]
\centering
\includegraphics[width=0.4\textwidth]{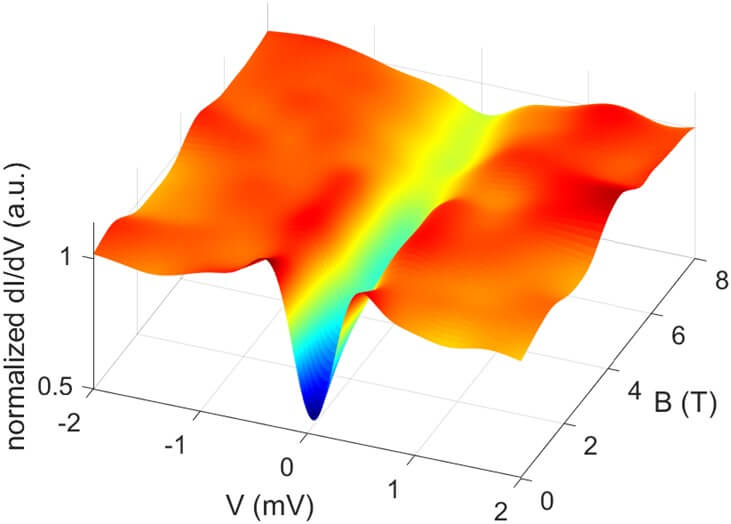}
\caption{3D plot of the magnetic field dependence of the tunneling
spectra of the phase with $\Delta(0) = 0.26$\,meV measured at $T = 0.5$\,K
in fields up to 8\,T.}
\label{fig:STM3}
\end{figure}

\section*{Upper critical field $B_{c2}$}
\begin{figure}[ht]
\centering
\includegraphics[width=0.35\textwidth]{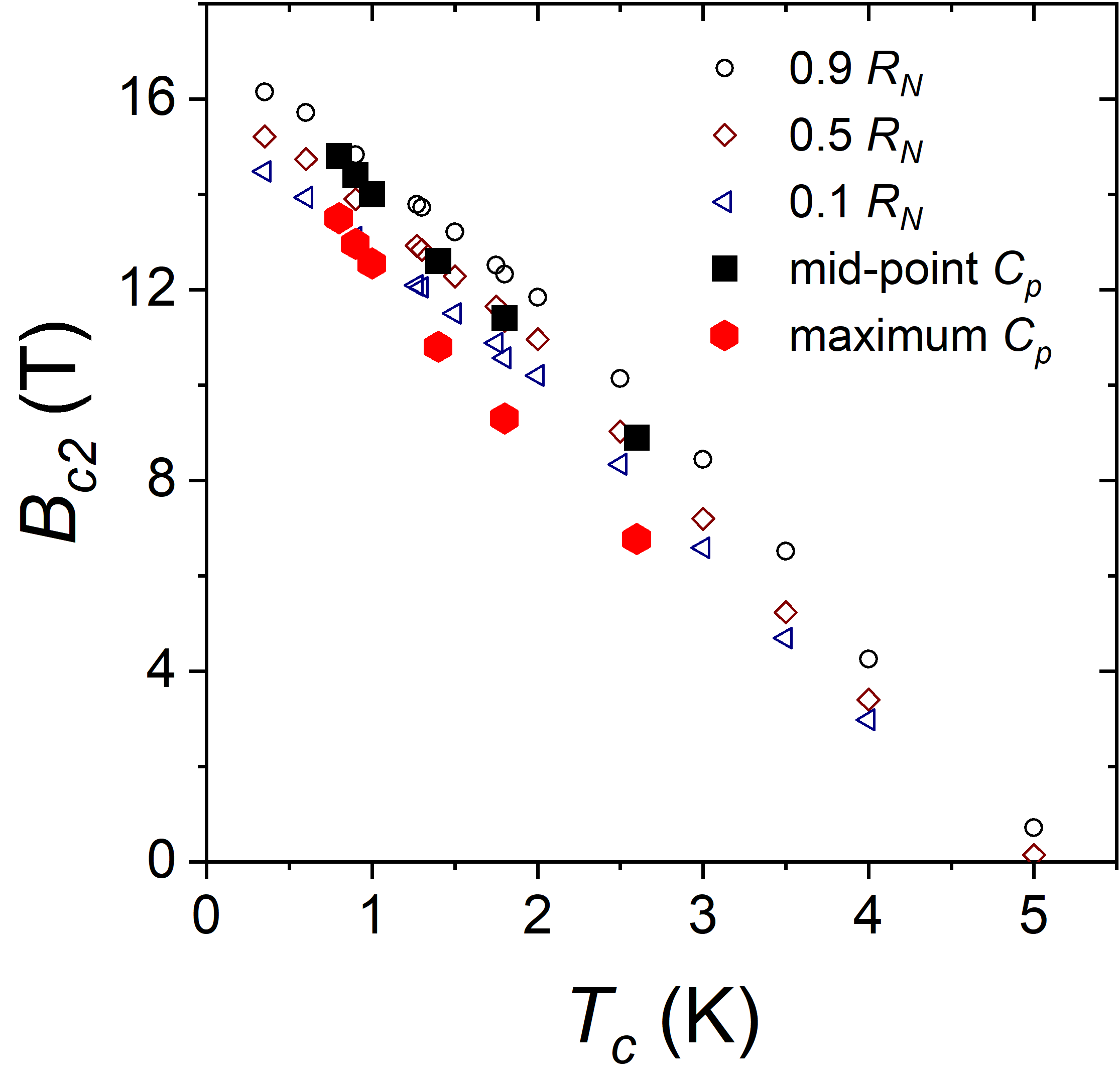}
\vspace{-3mm}
\caption{Upper critical magnetic field as a function of temperature
$B_{c2}(T)$ extracted from transport (open symbols) and heat-capacity
measurements. The critical fields from transport were determined from
the data shown in Fig.~\ref{fig:RvsH} of the Supplemental Material at 10\% (open
triangles), 50\% (open diamonds) and 90\% (open circles) of the normal
state resistance $R_\mathrm{N}$. The solid symbols indicate the $B_{c2}$
values determined from the magnetic field dependence of the heat capacity
at the mid-point (squares), and at the maximum (hexagons) of the
heat-capacity anomaly.}
\label{fig:Bc2}
\end{figure}

\vfill\eject

\section*{Two-gap models}
\begin{figure}[ht]
\centering
\includegraphics[width=0.32\textwidth]{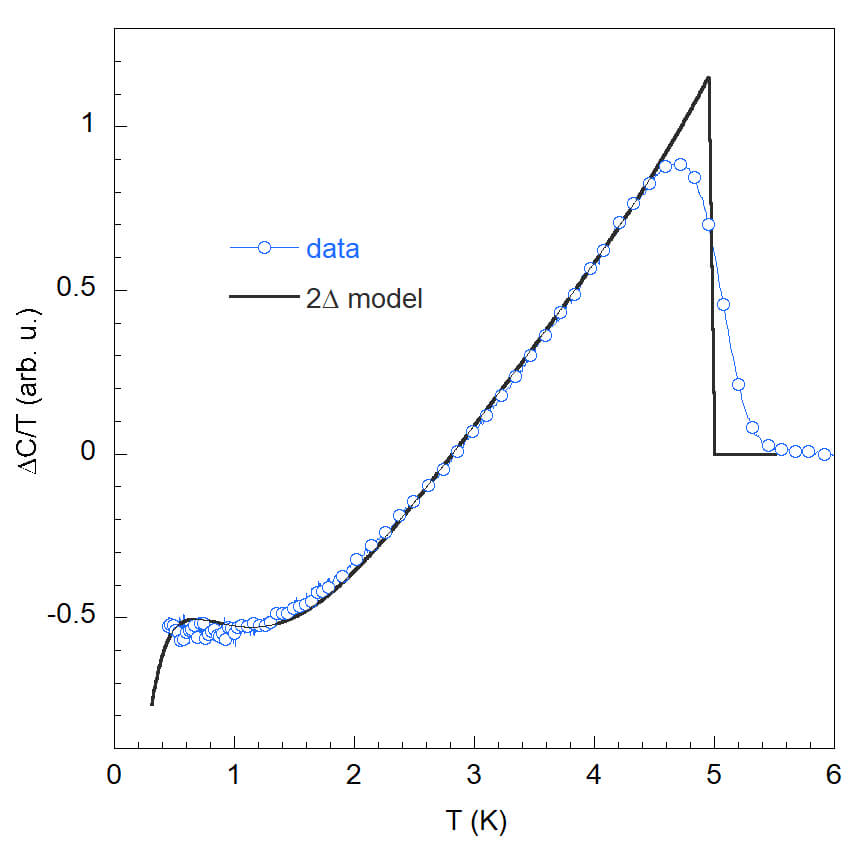}
\vspace{-3mm}
\caption{Heat capacity vs temperature. Alpha-model curve for two energy gaps with the parameters $2\Delta_1/k_\mathrm{B}T_c = 0.54$, $2\Delta_2/k_\mathrm{B}T_c = 4.3$ and equal weights of the two contributions (line), compared to the data (symbols).}
\label{fig:Cp for SM}
\end{figure}

\begin{figure}[ht]
\centering
\includegraphics[width=0.32\textwidth]{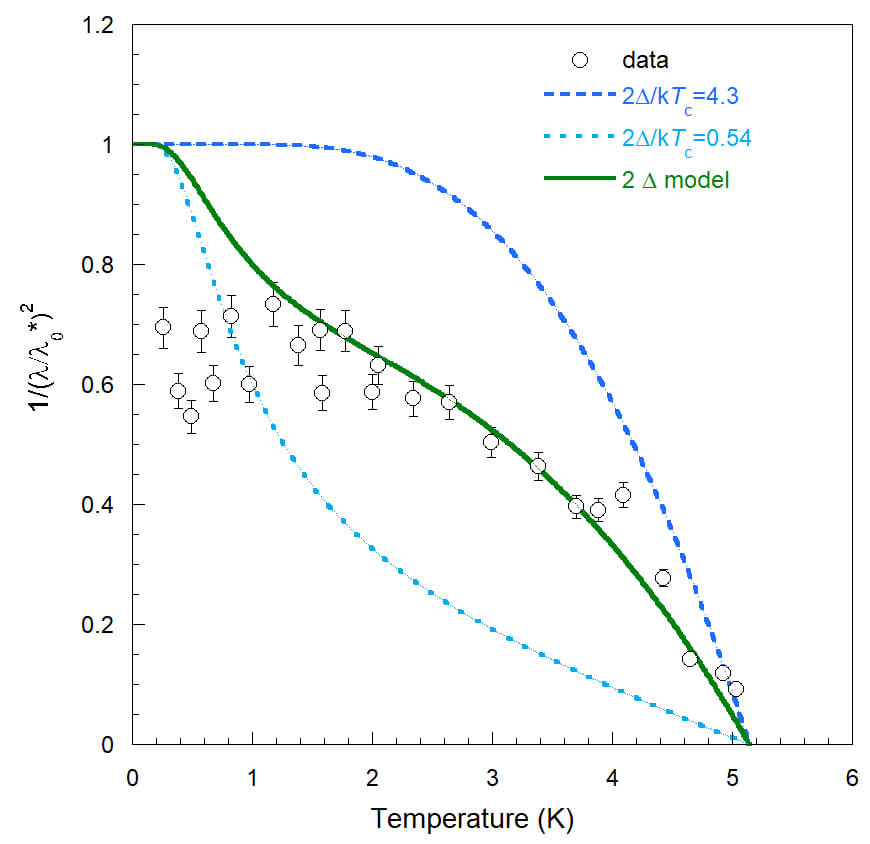}
\vspace{-5mm}
\caption{Normalized superfluid density of the sample (symbols), note that the data points were normalized in such a way as to obtain the best fit with the model (full line). Model curves of the superfluid density in the dirty limit following $\lambda^2(0)/\lambda^2(T) = [\Delta(T)/\Delta(0)]\tanh[\Delta(T)/(2k_\mathrm{B}T)]$ with the parameters $2\Delta_1/k_\mathrm{B}T_c = 0.54$ (dotted line), $2\Delta_2/k_\mathrm{B}T_c = 4.3$ (dashed line) and a sum of the two curves with equal contribution (full line).}
\label{fig:2D superfluid density}
\end{figure}
\begin{figure}[b!]
\centering
\includegraphics[width=0.32\textwidth]{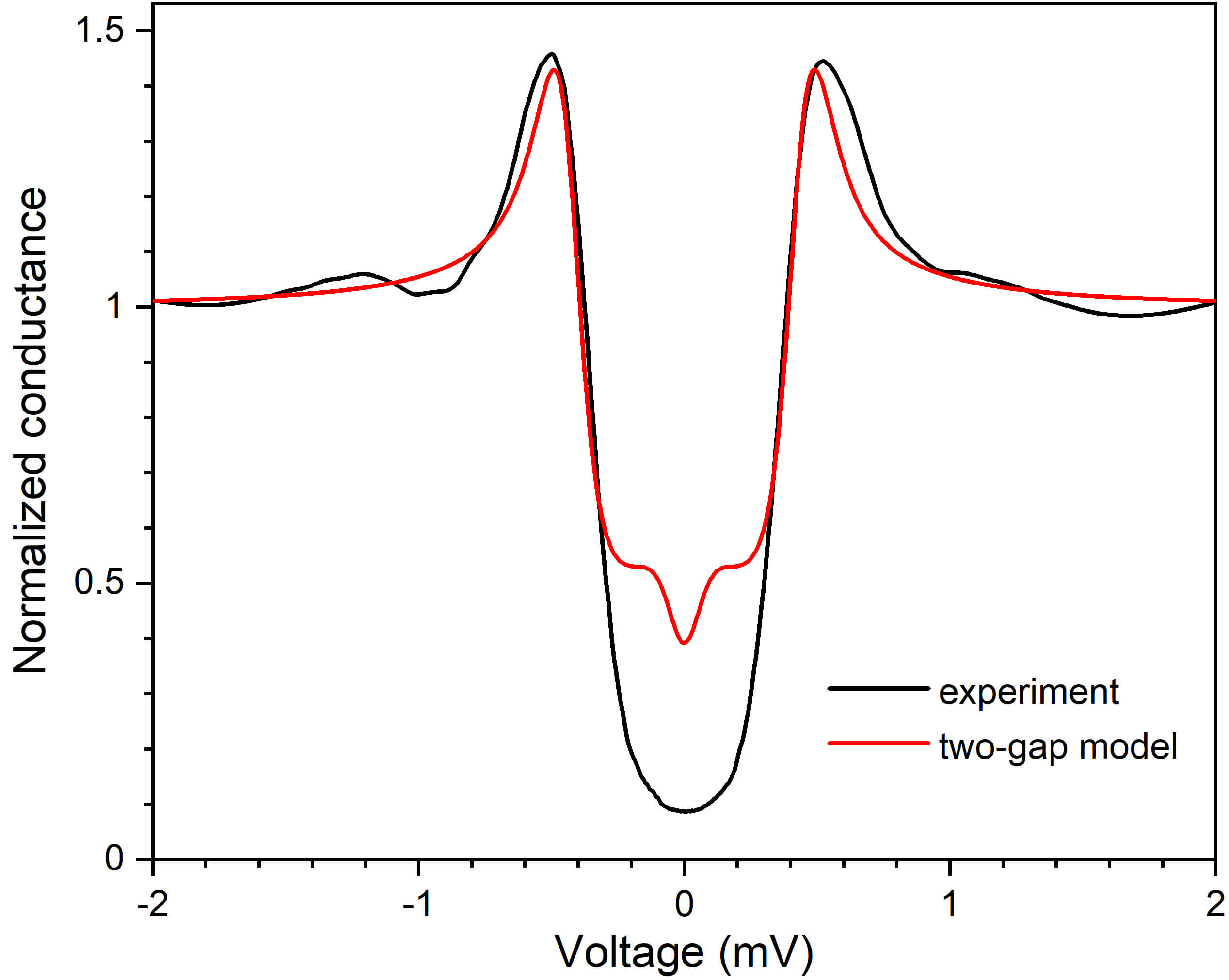}
\vspace{-3mm}
\caption{STM spectrum from Fig.2 of the paper (black line) and two-gap fit (red line) with the energy gap values $\Delta_{1}$ = 0.43 meV and $\Delta_{2}$ = 0.05 meV different by a factor of 8 and the same respective contribution $\alpha $= 0.5. The spectral smearings $\Gamma_1 $ and $\Gamma_2 $ were 10\% from the corresponding gap values.}
\label{fig:2D STM}
\end{figure}
\vfill\eject

\FloatBarrier
\makeatletter
\setlength{\@fptop}{0pt}
\makeatother

\begin{table*}[ht!]
\centering
\begin{tabular}{ccccccc}
\toprule
Atom & Wyckoff pos.\ & $x$ & $y$ & $z$ & Occupancy & $U_\mathrm{iso}$\\
\midrule
Ir1 & $4a$ & 0.19851 & 0.60825 & 0.19023 & 1.011 & 0.0153 \\
\hline
Ir2 & $4a$ & 0.34543 & 0.11003 & 0.11395 & 1.025 & 0.0039 \\
\hline
Ta & $4a$ & 0.00121 & 0.11984 & 0.00134 & 0.993 & 0.0041 \\
\hline
B1 & $4a$ & 0.01348 & 0.36993 & 0.35199 & 1.000 & 0.0130 \\
\hline
B2 & $4a$ & 0.19963 & 0.19194 & 0.35603 & 1.000 & 0.0110 \\
\bottomrule
\end{tabular}
\caption{Atomic coordinates and isotropic displacement parameters of TaIr$_2$B$_2$. The latter were refined for the Ir1, Ir2, and Ta atoms. The boron parameters were taken from those resulting from the single-crystal diffraction [doi.org/10.1002/adfm.202007960].}
\end{table*}
\FloatBarrier
\vfill\eject
\clearpage